\documentclass[useAMS,usenatbib]{mn2e}
\usepackage{graphics,epsfig,aas_macros}
\usepackage[normalem]{ulem}
\usepackage[dvipsnames]{color}
\usepackage[]{inputenc,amssymb}
\usepackage{amsmath}
\usepackage{color}
\usepackage{float}
\usepackage[toc,page]{appendix}

\def \be{\begin{equation}}
\def \ee{\end{equation}}
\def \bea{\begin{eqnarray}}
\def \eea{\end{eqnarray}}

\def\deg{\ifmmode^\circ\else$^\circ$\fi}
\def\pdeg{\ifmmode $\setbox0=\hbox{$^{\circ}$}\rlap{\hskip.11\wd0 .}$^{\circ}
          \else \setbox0=\hbox{$^{\circ}$}\rlap{\hskip.11\wd0 .}$^{\circ}$\fi}
\def\arcs{\ifmmode {^{\scriptstyle\prime\prime}}
          \else $^{\scriptstyle\prime\prime}$\fi}
\def\arcm{\ifmmode {^{\scriptstyle\prime}}
          \else $^{\scriptstyle\prime}$\fi}

\newcommand{\healpix}{\ensuremath{\tt HEALPix}}
\newcommand{\healpy}{\ensuremath{\tt Healpy}}

\title[Angular clustering of point sources at $150$~MHz]{Angular
  clustering of point sources at $150$~MHz in the TGSS survey} 

\author[Rana and Bagla]{
Sandeep Rana\thanks{E-Mail: sandeeprana@iisermohali.ac.in},
Jasjeet S. Bagla\thanks{E-Mail: jasjeet@iisermohali.ac.in}\\
Indian Institute of Science Education and Research Mohali,
Knowledge City, Sector 81, Sahibzada Ajit Singh Nagar, Punjab 140306,
India} 

\pubyear{2018}

\begin{document}

\label{firstpage}
\pagerange{\pageref{firstpage}--\pageref{lastpage}}

\maketitle

\begin{abstract}
  We study the angular clustering of point sources in The GMRT
  (Giant Meter Wave Telescope) Sky Survey (TGSS).
  The survey at $150$~MHz with $\delta > -53.5\deg$ has a sky
  coverage of $3.6\pi$ steradians, i.e., $90\%$ of the whole sky.
  We created subsamples by applying different total flux
  thresholds limit ($S_{flux} \gg 5 \sigma$) for good completeness and
  measured the angular correlation function $\omega(\theta)$ of point
  sources at large scales ($\geq 1\deg$). 
  We find that the amplitude of angular clustering is higher for
  brighter subsamples, this indicates that higher threshold flux
  samples are hosted by massive halos and cluster strongly: this
  conclusions is based on the assumption that the redshift
  distribution of sources does not change with flux and this is
  supported by models of radio sources. 
  We compare our results with other low-frequency studies of
  clustering of point sources and verify that the amplitude of
  clustering varies with the flux limit.
  We quantify this variation as a power law dependence of the
  amplitude of correlation function with the flux limit.
  This dependence can be used to estimate foreground contamination
  due to clustering of point sources for low frequency HI
  intensity mapping surveys for studying the epoch of
  reionisation.
\end{abstract}


\section{Introduction}

Low frequency observations of radio sources provide unique information
about the population of ultra-relativistic electrons in the
inter-stellar medium (ISM) of galaxies: synchrotron emission is the
primary radiative mechanism at these frequencies {\color{blue}
  \citep{1992ARAA..30..575C}}. 
Emission from ISM in galaxies and AGNs dominates over the
expected flux from neutral Hydrogen via the redshifted $21$~cm
radiation from the early Universe.
This constitutes a significant foreground that needs to be
characterized and removed in order to study the evolution of neutral
Hydrogen in the Universe. 
In particular these extragalactic foregrounds affects studies of the
epoch of reionisation (EoR) where the inter-galactic medium
transitions from being neutral to almost completely ionized 
{\color{blue} \citep{DiMatteo2004, Liu2009, Jelic08, Trott2012,
    Murray2017,2018arXiv180203060S,2018Natur.555...67B}}. 
Redshifted $21$~cm radiation from this epoch is likely to be observed
at wavelengths in the range $1.5-4$~m (frequency in the range
$75-200$~MHz). 
Therefore, a study of point sources and their clustering in this range
of frequencies is relevant not just from the perspective of studying
radio populations but also for its impact on EoR studies.
Studies of radio source clustering beyond angular correlation function
requires information about redshift, which is not available for an
overwhelming majority of sources at present. 

A number of studies have been carried out to quantify the faint source
population and their clustering at low frequencies.
A list of existing and ongoing radio surveys is presented in Table~1.

\begin{table*}
    \caption {Low frequency sky surveys.  This table enumerates
      sky surveys at low frequencies.  Their sky coverage,
      sensitivity, frequency and resolution are listed
      here.} \label{tab:surveys}   
    \begin{tabular}{ l  l  l  l } 
        \hline
        \hline
         Survey & Frequency(MHz) & Resolution & Noise (mJy~beam$^{-1}$)\\ 
        \hline
         VLSS({\color{blue} \cite{VLSS}}) & $74$  & $80^{\arcsec}$ & $100$  \\
         VLSSr({\color{blue} \cite{VLSSr}}) & $73.8$  & $75^{\arcsec}$ & $100$  \\
         8C({\color{blue} \cite{8C}}) & $38$  & $4.5^{\arcmin} \times 4.5^{\arcmin} \csc(\delta) $ & $200-300$ \\
         7C({\color{blue} \cite{7C}}) & $151$  & $70^{\arcsec} \times 70^{\arcsec} \csc(\delta) $ & $20$ \\
         MSSS-LBA({\color{blue} \cite{LOFAR_MSSS}}) & $ 30-78$ & $ \leq 150^{\arcsec} $ & $\leq 50$ \\
         MSSS-HBA{\color{blue} \cite{LOFAR_MSSS}} & $120-170$ & $ \leq
         120^{\arcsec} $ & $\leq 10-15$  \\
         TGSS {\color{blue} \cite{TGSSADR}} & $150$ & $25^{\arcsec}$ & $5$ \\
         GLEAM({\color{blue} \cite{MWA_GLEAM1}}) & $72-231$ & $100^{\arcsec}$ & $10$ \\
         LoTTS({\color{blue} \cite{LOFAR_2Mtr}}) & $120-168$ & $ 25^{\arcsec}$ & $0.5$ \\
        \hline
    \end{tabular}
\end{table*}

The Giant Metrewave Radio Telescope (GMRT) was used to survey the
radio sky at $150$~MHz between 2010 and 2012\footnote{Proposal for
the survey was made by Sandeep Sirothia, Nimisha Kantharia, Ishwara
Chandra and Gopal Krishna (GTAC Cycle 18).}.
Alternative Data Release (ADR1) {\color{blue} \citep{TGSSADR}} of the
TGSS survey contains a catalog of point sources.  
Here, TGSS data has been analyzed using the SPAM (Source Peeling and
Atmospheric Modeling) pipeline, which includes corrections for
direction-dependent ionospheric phase effects. 
Included in ADR1 are continuum stokes I images of $99.5\%$ of the
radio sky north of $\delta=-53^{\degr}$ ($3.6\pi$~sr, or $90\%$ of
the full sky) at a resolution of $25\arcsec \times 25\arcsec$ 
north of $\delta = 19\degr$ and $25\arcsec \times 25\arcsec /
\cos(\delta-19^{\degr})$ south of $\delta=19^{\degr}$, with a median
noise of $3.5$~mJy~beam$^{-1}$. 
ADR1 also provides a catalog of radio sources with coordinates, flux
density and sizes for $0.62$~million sources down to a $7~\sigma$ 
peak-to-noise
threshold\footnote{http://tgssadr.strw.leidenuniv.nl/doku.php}.    
The data analysis pipeline and data products are described in detail
in {\color{blue} \cite{TGSSADR}}.  

The survey sensitivity for about $80\%$ of the sky covered by TGSS is
$5$~mJy~beam$^{-1}$ or better (see figure~8 of {\color{blue}
  \cite{TGSSADR}}.   
The estimation of the TGSS confusion noise at $150$~MHz and with a
$25\arcsec$ beam ranges between $0.44$~mJy~beam$^{-1}$ and $2.5$~mJy
beam$^{-1}$ for most of the sky. 
The TGSS point source survey has $50 \%$ completeness at $25$~mJy (or
$7\sigma$ for point sources, with $\sigma$ being the median survey
noise of $3.5$~mJy~beam$^{-1}$).
For more detail see {\color{blue} \cite{TGSSADR}}.
We choose to work with subsets with peak flux thresholds $> 32$~mJy
cutoff to ensure better completeness. 

%%%%%%%%%%%%%%%%%%%%%%%%%%%%%%%%%%%%%%%%%%%%%%%%%%%%%%%%%%%%%%%%%%%%%%%%

\section{Analysis and results}

\subsection{Survey Selection}

Our main aim here is to do study clustering of point sources. 
We require a sample that is homogeneous and complete for this
purpose. 
The ADR1 data provides the peak flux, source flux and noise on the
individual sources.
We created a pixelised all-sky map for all sources in TGSS ADR1
catalog with $Nside=1024$. 
This corresponds to mean spacing of $0.057\deg$, using \healpy~the
python version of \healpix\footnote{http://healpix.sourceforge.net}
{\color{blue}\citep{Healpix}}. 
The mean spacing between pixels is much larger than the nominal
resolution of $0.0069\deg$ but much smaller than the primary beam.   
We mask out the region $|b| \leq 10\deg$ in order to avoid
contamination from galactic sources. 
We also mask out regions of the sky that cannot be observed using the
GMRT, i.e., $\delta \leq -53\deg$. 
We then mask all pixels with noise level of $>4$~mJy~beam$^{-1}$.
Sources in the remaining pixels can potentially be used for further
studies.
A binary mask that allows the remaining pixels is also used for
generating random catalogs. 
To study clustering of radio sources, we now create a catalog for
sources with peak flux $> 32$~mJy~beam$^{-1}$ after imposing the same
binary mask. 
We consider this as the master catalog from where subsets are
generated for clustering analysis.
We created different source population subsets with flux threshold
using total source flux of $\geqslant 50$~mJy, $60$~mJy, $100$~mJy and
$200$~mJy respectively.
The full ADR1 catalog is $90\%$ complete at total source flux of
$60$~mJy, with our cut of $4$~mJy~beam$^{-1}$, we expect better
completeness at lower flux thresholds. 
In order to ensure completeness, we do not use the catalog with a peak
flux of $32$~mJy~beam$^{-1}$ but use higher flux thresholds instead. 
The total number of source in each subset are $267752$, $239993$,
$163654$, and, $87751$ respectively.
$S \geqslant 60$~mJy at $150$~MHz corresponds to NVSS $S\geqslant
10$~mJy, according to the typical spectral index relation
{\color{blue} \citep{Tiwari, TGSSADR}}: 
\begin{equation}
\alpha_{obs} = \frac{(log S_{TGSS} - log S_{NVSS})}{(log \nu_{NVSS} - 
  log \nu_{TGSS})}  
\end{equation}
Where $\nu_{NVSS}=1.4$~GHz, $\nu_{TGSS}=150$~MHz and $S_{TGSS}$ and
$S_{NVSS}$, respectively, at the flux densities measured by the TGSS
and NVSS for sources common in the two catalogs. 
We used $\alpha_{obs} = 0.76$ for conversion (see figure 2 of
{\color{blue} \cite{Tiwari}} for details). 

Two sky maps are included as online only supplementary material.
One map shows the regions with {\sl rms} noise below
$4$~mJy~beam$^{-1}$ and the second map shows all the sources in the
$50$~mJy catalog.

\begin{figure}
	\begin{center}
        \includegraphics[width=7.5cm,angle=0.0]{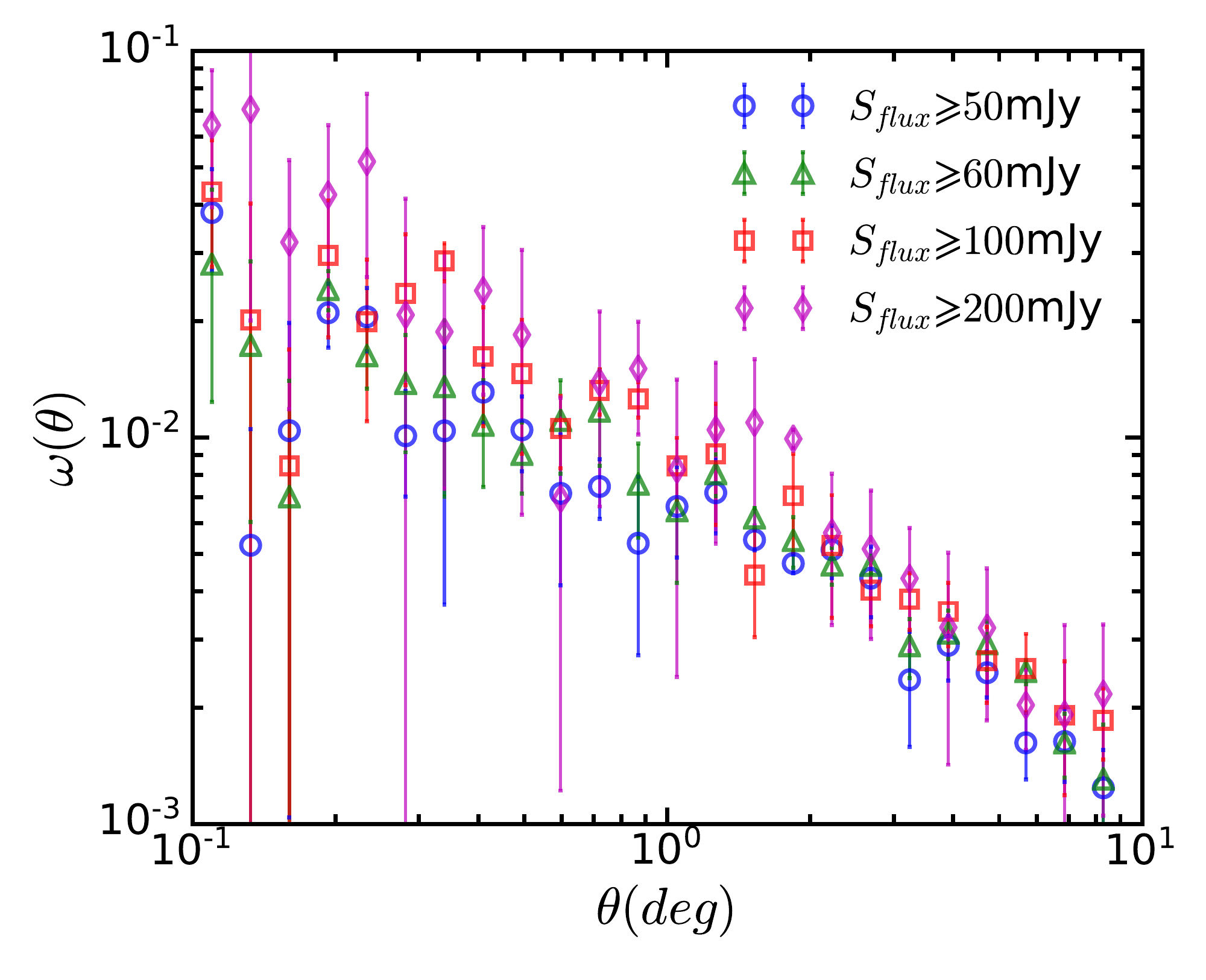}
        \caption{Angular correlation function for the four subsets.
          We see that there is a break in the shape at an angle of
          $1^\circ$ with a power law decline at larger scales and a
          gentler variation at smaller scales.  The amplitude of
          angular correlation is larger for subsamples with a high
          flux threshold, i.e., brighter sources cluster more
          strongly.} 
        \label{2}
	\end{center}
\end{figure}

\subsection{Angular Correlation Function}

The angular correlation function $\omega(\theta)$ describes the
clustering of sources on the sky. 
It is a measure of the excess number of neighboring sources at a
separation $\theta$ on an average, where the excess is measured over a
random distribution of sources with the same number density
{\color{blue}\citep{1980lssu.book.....P}}.  
We calculate $\omega(\theta)$ using the Landy-Szalay 
{\color{blue}\citep{LS}} estimator which takes the edge corrections 
{\color{blue} \citep{2ptCorr_Compare}} into account.
The angular correlation function is defined in terms of pair counts in
the data and the random catalog:  
\begin{equation}
      \omega(\theta) = \frac{N_{r}(N_{r}+1)DD}{N_{d}(N_{d}+1)RR} -
      N_{d}(N_{r}+1)\frac{DR}{RR} + 1 
\end{equation}
Here, $DD$ denotes the count of pairs in the data at angular separation
$\theta$, and $N_{d}$ is the total number of objects in the data
considered in the analysis.
Similarly, $RR$ is the averaged pair count over a catalog of uniformly
distributed points covering the same survey area, $DR$ is the
data-random cross pair count, and $N_{r}$ denotes the number of points
in the random catalog. 

We created a random catalog for clustering analysis using the
  survey selection function as described in the previous section.  

  To compute $DD$, $RR$ and $DR$ efficiently, we use the publicly
  available KD-tree routines (Scikit-learn
  {\color{blue}\citep{sk}}) which are an implementation of spatial
  algorithms such as kd-tree for fast
  nearest neighbor search {\color{blue} \citep{Tree,Tree1}}.
  We used the Jackknife re-sampling method  {\color{blue}
    \citep{bootstrap, bootstrap1}}, with $200$ subsamples drawn 
  randomly from the flux limited data to estimate errors in
  $\omega(\theta)$.
  In this work, all the plots were produced with MATPLOTLIB
 {\color{blue} \cite{Hunter:2007}} and frequently used scientific
 libraries such as NUMPY and SCIPY {\color{blue}\citep{Numpy,Scipy}}.

%%%%%%%%%%%%%%%%%%%

\begin{figure*}
	\begin{center}
        \includegraphics[width=12cm,angle=0.0]{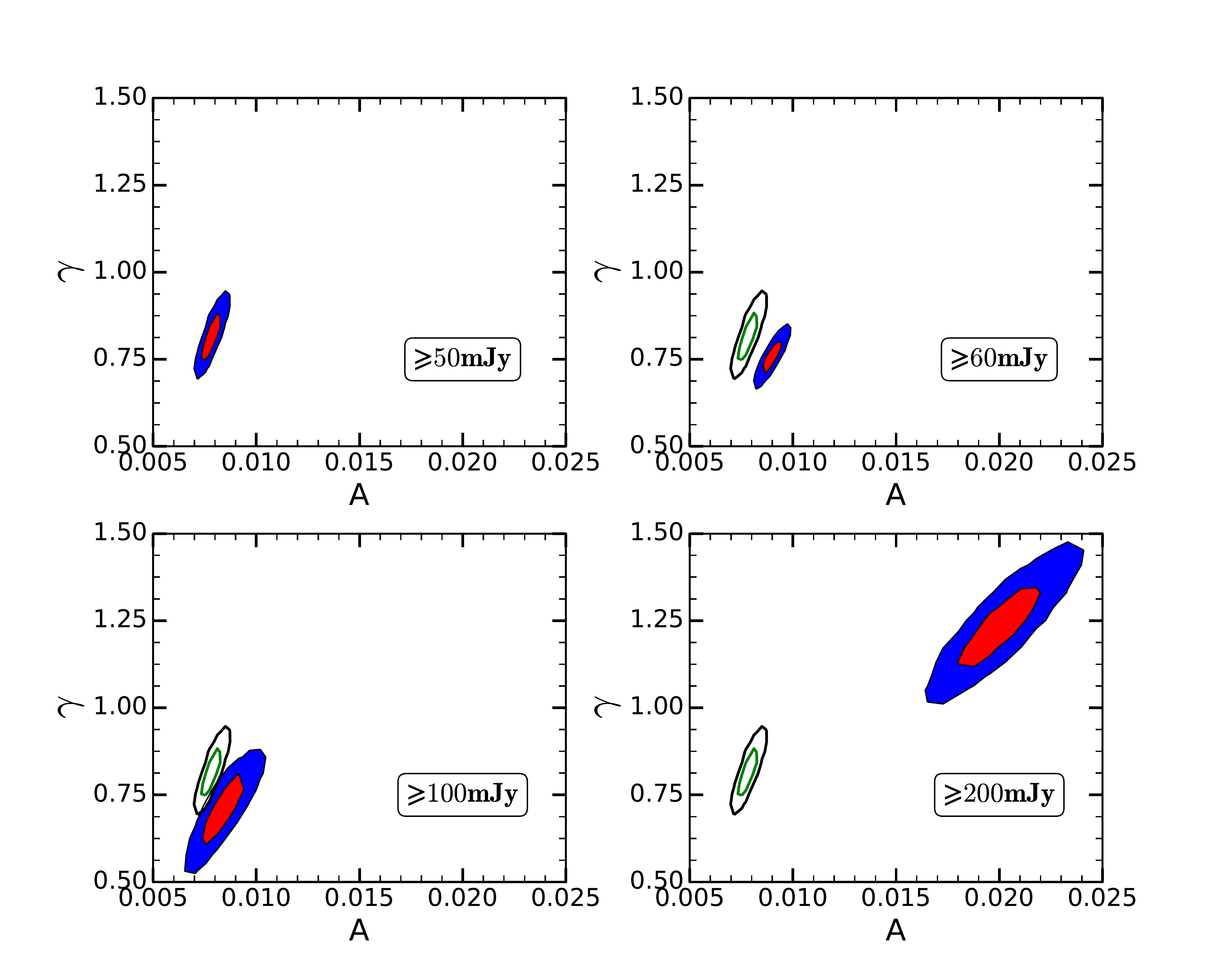}
		\caption {Posterior probability distribution for $A$
                  and $\gamma$ using angular correlation data for
                  different flux cut-off values.  This is shown for
                  the four subsets.  $1\sigma$ and $2\sigma$
                  contours are shown here.  The outline of contours
                  for the $50$~mJy subset is shown in all the panels
                  for reference.  Note that the power law is fit only
                  at scales $\theta \geq 1^{\circ}$.}  
		\label{3}
	\end{center}
\end{figure*}

\begin{figure*}
	\begin{center}
        \includegraphics[width=12cm]{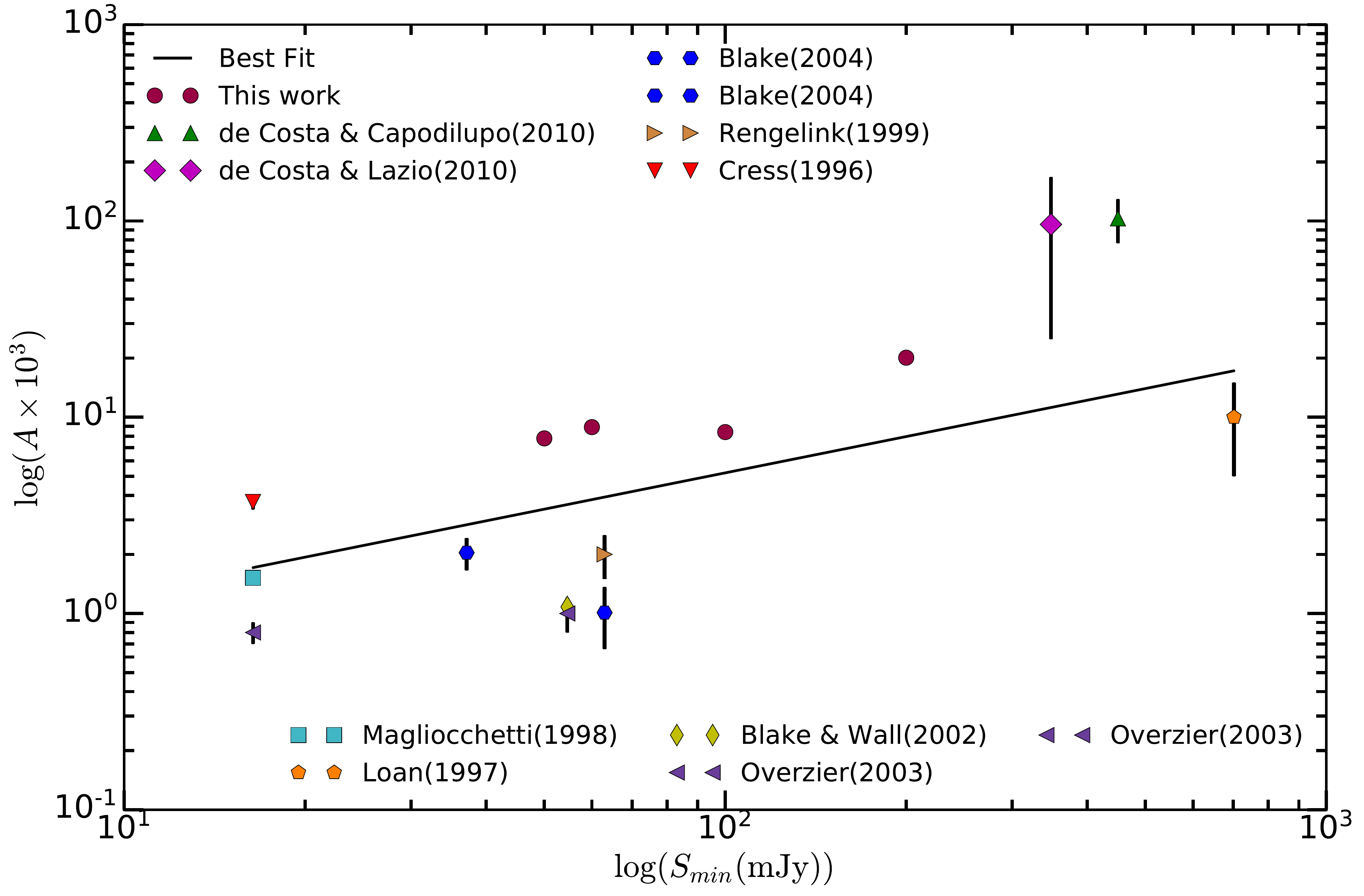}
        \caption {We show the amplitude of correlation
                  function as a function of the scaled flux threshold
                  from different studies. See text for details of the
                  scaling procedure. We find a trend of a higher
                  correlation amplitude for a higher flux threshold.
                  The line plotted in the top panel is the
                  best fit power law for these data points of the form
                $(a \times 10^{-3} +b~log \left(S_{min}/\left(1\,\mathrm{mJy}\right)\right)) $. The values are: $a=-0.51 \pm 0.59$ and $b=0.61 \pm 0.33$, as compared to the values reported earlier: $a = 0.16 \pm 0.18$ and $b=0.35 \pm 0.13$.}
		\label{4}
	\end{center}
\end{figure*}

\begin{center}
\begin{table*}
    \caption {Best fit value of $\omega(\theta)$ at low frequencies
      from current study and earlier published work.} \label{tab:bfwtheta}  
    \begin{tabular}{l l l l l l l l} 
        \hline
        \hline
         Survey & Ref & $\nu$~(MHz)  & $S_{lim}(mJy)$ & $A$  &
                                                                    $\gamma$
      & Number of & Scaled flux \\
      & & Sources & & ($\times 10^{-3}$) & & &  at $150$~MHz (mJy) \\
        \hline
        \hline
         TGSS &   This Work & 150  & $\geqslant 50$  & $7.8 \pm 0.4$ &
                                                                       $0.821
                                                                       \pm
                                                                       0.072$
      & 267752 & $50$ \\
         TGSS &   This Work & 150  & $\geqslant 60$  & $8.9 \pm 0.5$ &
                                                                       $0.760 \pm 0.053$  & 239993 & $60$\\
         TGSS &   This Work & 150  & $\geqslant 100$ & $8.4 \pm 0.1$ &
                                                                       $0.716
                                                                       \pm
                                                                       0.104$
      & 163654 & $100$ \\
         TGSS &   This Work & 150  & $\geqslant 200$ & $20.1\pm 0.2$ &
                                                                       $1.24
                                                                       \pm
                                                                       0.132$
      & 87751 & $200$ \\
        \hline
         VLSS &  {\color{blue}\cite{deCosta2010_74MHz}} & 74  & 770 &
                                                                      $103
                                                                      \pm
                                                                      26$
                                                                  &
                                                                    $1.21
                                                                    \pm
                                                                    0.35$&
                                                                           68311 & $450.1$ \\ 
        \hline
        MIYUN &  {\color{blue}\cite{deCosta2010}}       & 232 & 250 &
                                                                      $96
                                                                      \pm
                                                                      71$
                                                                  &
                                                                    $1.12
                                                                    \pm
                                                                    0.11$&
                                                                           34426
                          & $348.2$ \\
        \hline
        PMN   &   {\color{blue}\cite{Loan97}}        &
                                                       $4.85\times10^{3}$
                                     & 50 & $10.0\pm5.0$ & 1.8& 77856
                          & $701.9$ \\
        \hline
        WENSS &   {\color{blue}\cite{WENSS1999}}     & 325  & 35 &
                                                                   $2.0\pm0.5$
                                                                  &
                                                                    0.8 & 86461 & $62.9$ \\
        \hline
        WENNS &   {\color{blue}\cite{Blake2004}}     & 325  & 35 &
                                                                   $1.01\pm0.35$
                                                                  &
                                                                    $1.22\pm0.33$&86461
                          & $62.9$ \\
        \hline
        SUMMS &   {\color{blue}\cite{Blake2004}}     & 843  & 10 &
                                                                   $2.04\pm0.38$
                                                                  &
                                                                    $1.24\pm0.16$&68373 & $37.1$ \\
        \hline
        FIRST &   {\color{blue}\cite{Cress1996}}     & 1400 & 3  &
                                                                   $3.7\pm0.3$
                                                                  &
                                                                    $1.06
                                                                    \pm0.03$
      & 109873 & $16.4$ \\
        \hline
        FIRST &   {\color{blue}\cite{Magliocchetti}} & 1400 & 3  &
                                                                   $1.52
                                                                   \pm0.06$
                                                                  &
                                                                    $1.68\pm0.07$
      & 86074 & $16.4$ \\
        \hline
        NVSS  &   {\color{blue}\cite{Blake2002}}      & 1400 & 10 &
                                                                    $1.08
                                                                    \pm0.09$
                                                                  &
                                                                    $0.83\pm0.05$ & 522341 & $54.6$\\
        \hline
        NVSS  &   {\color{blue}\cite{Overzier2003}}   & 1400 & 3  &
                                                                    $0.8
                                                                    \pm0.1$
                                                                  &
                                                                    $0.8
                                                                    $
      & 210530 & $16.4$ \\
        NVSS  &   {\color{blue}\cite{Overzier2003}}   & 1400 & 10  &
                                                                     $1.0
                                                                     \pm0.2$
                                                                  &
                                                                    $0.8 $ & 433951 & $54.6$ \\
        \hline
        \hline
    \end{tabular}
\end{table*}
\end{center}

We find that the amplitude of the angular correlation function varies
monotonically with the flux cut-off at all scales, as shown in
Figure~\ref{2}.
As the flux threshold is increased, the amplitude of angular
correlation function at a fixed angle increases. 
This is consistent with findings in earlier studies, e.g.,
{\color{blue}\citep{1991MNRAS.253..307P,WENSS1999,2017MNRAS.464.3271M}}.
This may arise from the known correlation between the star formation
rate and the stellar mass of galaxies, e.g., {\color{blue}
  \citep{2013MNRAS.434..451L}}.
It is well known that stellar mass and halo masses of galaxies are
correlated and higher mass halos are more strongly biased.
This may also result from a higher prevalence of AGN in more massive
galaxies.
While not all high mass galaxies are AGNs, more powerful AGNs are to
be found mostly in high mass galaxies and hence are likely to be more
strongly biased.
Qualitatively we expect this to be true for radio loud AGNs as well
 {\color{blue}\citep{WENSS1999,Overzier2003}}.  
Given the flux range, the majority of sources in our sample are
expected to be AGNs.

To quantify the shape of the angular correlation function, we assume a 
power law of the form: $A \theta^{-\gamma }$ {\color{blue}
  {\color{blue}\citep{Cress1996,WENSS1999, Magliocchetti,
      Blake2002,Overzier2003, Blake2004}}}. 
Here $A$ is the amplitude, $\theta$ is the angle in degrees and
$\gamma$ is the power law index.
We estimate the posterior probability distribution alongside best
fit for amplitude $A$ and power law index $\gamma$. 
We use the {\sl emcee} package {\color{blue} \citep{emcee}}, which
is a Python implementation of MCMC sampling for this estimation.
We fit the power law to data at angular scales larger than a
degree.
This allows us to avoid modeling a change of slope seen in some of
the subsets at the scale of a degree: the angular correlation function
is shallower at smaller angular scales and its slope appears to vary
strongly with the flux cutoff. 
Further, most EoR experiments are sensitive to larger angular scales
and hence we focus on these scales\footnote{For reference, the
  estimated slope and amplitude if we work with the range $0.1^\circ
  \leq \theta \leq 10^\circ$ is shown in a separate figure available
  online in supplementary material.  As can be seen, the slope for the
full range of scales is much smaller for all the sets, as is the
amplitude.  This shows that the slope at smaller angular scales is
gentler for all the subsamples.}

Figure~\ref{3} shows the $1 \sigma$ and $2 \sigma$ confidence 
intervals for the subsamples with different flux cut-off, plotted in 
the $A-\gamma$ plane.
We see that as the flux cut-off increases, the dominant effect is an
increase in $A$. 
The preferred range of $\gamma$ remains within $1\sigma$ region for
the $50$~mJy catalog and it does not show any strong, systematic
evolution with the flux cut-off, except for the brightest subset.
Such a behavior may be expected if most of the sources belong to the
same class.

We compare our results with other studies of radio source clustering. 
In Table \ref{2} we have listed clustering studies of radio sources in
surveys carried out at different frequencies
{\color{blue}\citep{deCosta2010}}.  
We find that if we consider surveys with a comparable flux cut-off, as
determined by the average spectral index, the amplitude of clustering
is comparable within uncertainty. 
This is illustrated in Figure~\ref{4}.
This figure shows the variation of the amplitude of angular clustering
as a function of the scaled flux threshold in different samples
assuming a spectral index $0.76$.
We find that there is a clear trend such that increasing the scaled
flux threshold leads to a higher amplitude of clustering. 
However, we see some scatter around this trend, unlike figure~1 of 
{\color{blue} Rengelink et al. 1999}.
The source of the scatter is unclear and it may be due to sample
variance or the range of angular scales over which a reliable
estimation of the angular correlation function can be done in a given
sample {\color{blue}\citep{2000MNRAS.314..546M,Blake2002,Overzier2003}}.
Angular resolution can play a role at small angular scales while the
extent of the survey on the sky and the geometry can effect the
largest scales up to which we can get a reliable estimate of the
angular correlation function and not be drowned out by sample/cosmic
variance. 
Another aspect is the use of a single spectral index for scaling
observations: variations in the spectral index in different samples
can also introduce some scatter.

\begin{figure}
	\begin{center}
        \includegraphics[width=7.5cm]{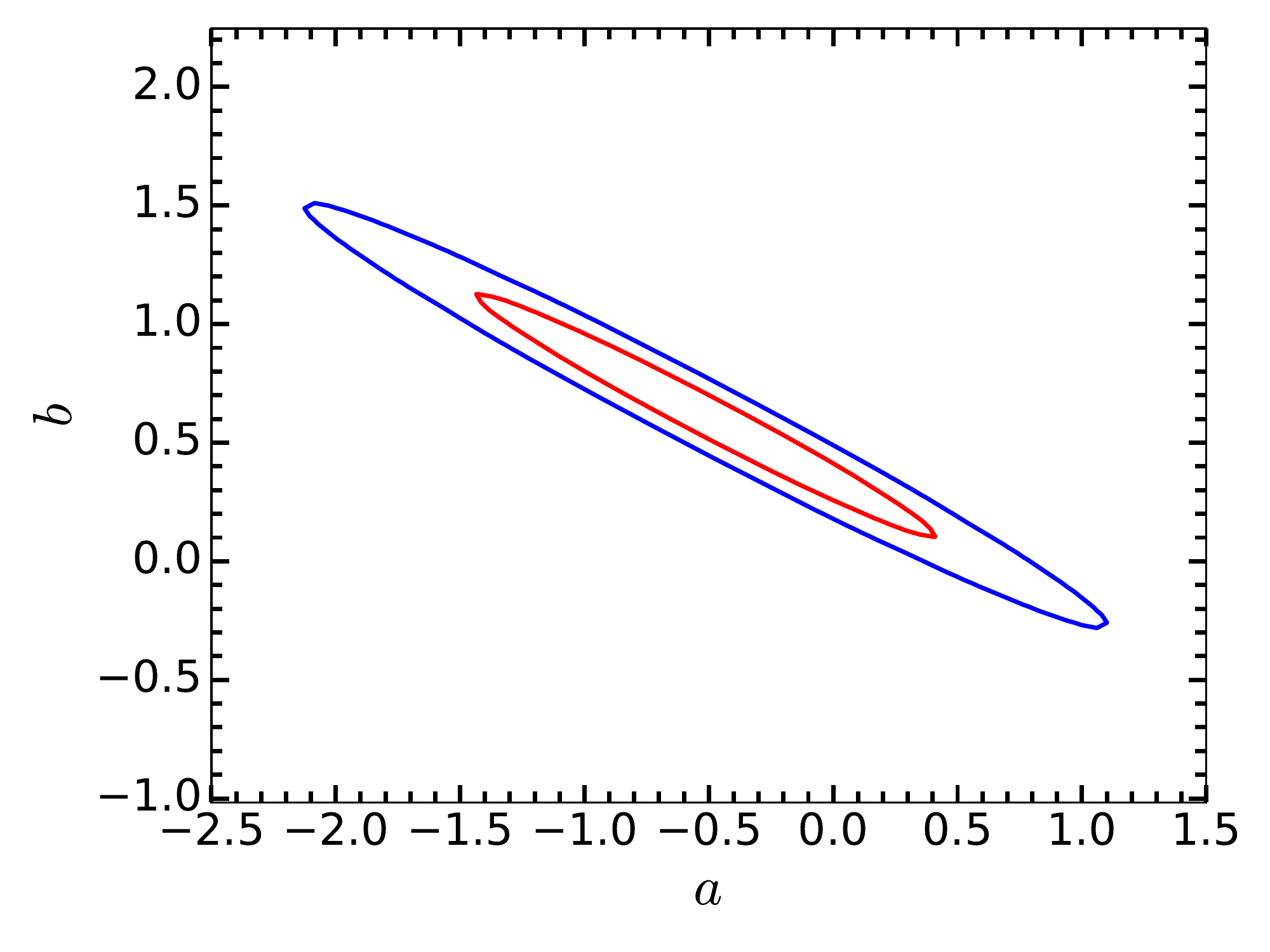}
        \caption {We show confidence level intervals
                of the power law fit in the amplitude ($a$) and index
                ($b$) space.}
		\label{4a}
	\end{center}
\end{figure}

The line plotted in the top panel of Figure~\ref{4} is the best fit
power law for these data points and has the form $a \times 10^{-3}
  +~b~log\left(S_{min}/\left(1\,\mathrm{mJy}\right)\right)$.
The fit is driven by points with smaller error bars.  
Figure~\ref{4a} shows the confidence levels of the power law
fit. 
There is no significant variation in the index $\gamma$ of the angular
correlation function with flux limit in our sample at lower flux
levels.
Thus, the fitted variation of the amplitude of angular clustering can
be used for simulations of point source foregrounds at and around
$150$~MHz.
Indeed, this allows for an approach that is independent of the
modeling of different types of radio sources {\color{blue}
  \citep{DiMatteo2004, Jelic08, Liu2009, Trott2012, Murray2017}}.

Clustering has been analyzed for deeper surveys, e.g.,
{\color{blue} \citep{2003MNRAS.339..695W}}, however such surveys so
far have covered 
a smaller area in the sky and hence are not appropriate for comparison
with wide field surveys, except perhaps for studying clustering at
sub-degree scales.

%%%%%%%%%%%%%%%%%%%

\section{Discussion}

We discuss our results in the context of source populations and the
implications for foreground characterization.

A number of models for radio sources have been developed
{\color{blue}\citep{2010MNRAS.404..532M,2017ApJ...842...95M,2017MNRAS.469.1912B,2018IAUS..333..175P,2019MNRAS.482....2B}}. 
While early models were largely phenomenological, improvement in
multi-wavelength observations have led to development of more
sophisticated and realistic models.  
These models suggest that at low frequencies and at flux ranges of
interest, the data sets are dominated by radio loud AGNs.
Indeed, at $150$~MHz, steep spectrum sources account for nearly the
entire source population at fluxes relevant for our dataset
{\color{blue}\citep{2010MNRAS.404..532M,2017ApJ...842...95M}}.
As we go to fainter fluxes, we probe sources at a slightly higher
redshifts though the shift is very small compared with the range of
redshifts over which the sources are distributed
{\color{blue}\citep{2010MNRAS.404..532M,2017ApJ...842...95M}}.  
The fact that we probe primarily one class of sources is the likely 
reason for the observed variation of clustering amplitude with flux as
in  {\color{blue}\citet{Overzier2003}}.
The observed source population in TGSS consists mostly of AGNs and as
brighter AGNs reside in more massive halos, we expect these to be more
strongly biased  {\color{blue}\citep{Overzier2003,2017MNRAS.464.3271M}}.  
We are analyzing this is detail by combining source population models
with models for evolution of bias.
Results of the study will be reported in a forthcoming manuscript.
Preliminary analysis suggests that at lower fluxes, below $1$~mJy
at $150$~MHz, star forming galaxies begin to contribute significantly
and the variation of clustering with flux should change character and
deviate from the power law behavior seen
here {\color{blue}\citep{2003MNRAS.339..695W}}. 

The quantitative analysis of point source clustering can be used as an
input for foreground removal
{\color{blue}\citep{Murray2017,2018IAUS..333..199M}}. 
Essentially, the foreground removal methods result in separation of
different contributions and an independent assessment of the
foregrounds can be used as a calibrator for the foreground removal
process.
Bright point sources can be removed explicitly and hence a statistical
approach is required only for the fainter sources.
Smaller amplitude of angular clustering for fainter sources is hence
significant in that foregrounds due to clustering of faint sources may
be less important than estimated from bright samples.
We expect that EoR surveys should be able to remove sources brighter
than about $0.1$~mJy and hence it is important to estimate angular
clustering for fainter radio source populations.
However, as we gradually work towards this sensitivity, it is
essential to work with the data sets available and refine algorithms.
Studies of clustering of sources over a wide range in fluxes allow us
to validate models for source populations and their clustering: this
is essential if we are to understand and work around foregrounds for
the sensitive EoR surveys.

\section{Summary}

In this paper we have studied the angular clustering of radio sources in
the TGSS survey using the catalogs derived in the alternative data
release.
We have defined our main sample and sub samples using the rms noise
and peak flux.
We have studied angular clustering of these sources our main
results are:
\begin{enumerate}
\item
  The angular correlation function is a power law at scales
  larger than a degree: correlation drops rapidly at larger angular
  separations. 
\item
  The angular correlation at scales
  smaller than a degree has a weaker dependence on scale as compared
  to larger scales.
  This is illustrated in a figure available online as supplementary
  material. 
\item
  The slope of the angular correlation function shows little variation
  with the flux of sources.
\item
  The amplitude of the power law increases monotonically with the peak
  flux of sources.  This is consistent with earlier studies, 
  {\color{blue} \citep{1991MNRAS.253..307P,WENSS1999,2003MNRAS.339..695W,Overzier2003,2017MNRAS.464.3271M}}. 
\item
  We have compared our results with other studies of angular clustering
  of radio sources.
  We show that assuming a typical spectral index of $\alpha=0.76$, the
  amplitude of angular clustering is insensitive to the frequency at
  which the sources are observed and selected. 
  This is in agreement with {\color{blue}\cite{WENSS1999}}.
\item
  We provide a fit to the variation of the amplitude of clustering
  with the flux cutoff at $150$~MHz.
  This is potentially useful for modeling of point source foregrounds
  for EoR studies. 
\end{enumerate}

\section*{Acknowledgments}

Authors thank Manuela Magliocchetti and Nirupam Roy for useful
comments and suggestions.
We also thank Gianfranco de Zotti for clarifications on models of radio
sources.
We thank the anonymous reviewer for detailed comments.
Authors thank Thilo Max Siewert for pointing out an error in Figure~3 and Figure~4 in the published version.
The authors acknowledge the use of the HPC facility at IISER Mohali
for this work.
This research has made use of NASA's Astrophysics Data
System Bibliographic Services.

\bibliographystyle{mnras}
%\iffalse

%\fi

%\bsp  
\label{lastpage}
\end{document}